\documentclass[conference]{IEEEtran}
\IEEEoverridecommandlockouts
\usepackage{cite}
\usepackage{amsmath,amssymb,amsfonts}
\usepackage{algorithmic}
\usepackage{graphicx}
\usepackage{textcomp}
\usepackage[flushleft]{threeparttable}
\usepackage{url}
\usepackage[dvipsnames]{xcolor}
\usepackage{makecell}
\newcommand{\RNum}[1]{\uppercase\expandafter{\romannumeral #1\relax}}
\usepackage[numbers,sort&compress]{natbib}
\def\BibTeX{{\rm B\kern-.05em{\sc i\kern-.025em b}\kern-.08em
    T\kern-.1667em\lower.7ex\hbox{E}\kern-.125emX}}

\begin{document}
\bstctlcite{IEEEexample:BSTcontrol}
\title{Integration of  Renewable Generators in Synthetic Electric Grids for Dynamic Analysis\\

}

\author{\IEEEauthorblockN{Yijing Liu, Zeyu Mao, Hanyue Li, Komal~S~Shetye and Thomas J. Overbye}
\IEEEauthorblockA{Department of Electrical and Computer Engineering}Texas A$\&$M University,
College Station, TX, USA 77843\\
Email: $\{$yiji21, zeyumao2, hanyueli, shetye, overbye$\}$@tamu.edu}

\maketitle

\begin{abstract}
%Synthetic electric grids that are fictitious models have the capability of capturing the important features of actual power grids without any confidentiality concerns. 
%Synthetic electric grids provide engineers close-to-real models to study without any confidentiality concerns.
%Previous work has presented an approach to extend synthetic grids for dynamics studies. With the rapid growth of solar and wind energy world-wide in the past decade, there has been an urgent need to include dynamic models for these renewable generators. 
This paper presents a method to better integrate dynamic models for renewable resources into synthetic electric grids. An automated dynamic models assignment process is proposed for wind and solar generators. A realistic composition ratio for different types of wind turbine generators (WTG) is assigned to each wind generator. Statistics summarized from real electric grid data form the bases in assigning proper models with reasonable parameters to each WTG. A similar process is used to assign appropriate models and parameters to each photovoltaic (PV) generator. 
%Several model tuning methods are utilized to validate the dynamic response of the renewable generators. 
%5This model and parameter template assignment methodology is illustrated and validated in a 2000-bus test case. 
Multiple control strategies of the renewable resources are considered and tested in case studies. Two large-scale synthetic network test cases are used as examples of modeling the dynamics of renewable generators. 
%The added dynamic models for the renewable generators have been extensively tested for multiple control strategies. 
Several transient stability metrics are adopted to assess the stability level after being subject to N-1 contingency event. Representative contingency events are given to demonstrate the performance of the synthetic renewable generator models. 
%A representative N-2 contingency event is given to demonstrate renewable generators' dynamic models have impact on system stability.
 \end{abstract}


\section{Introduction}
Appropriate power system dynamic models are crucial for system engineers to get more realistic and insightful transient stability analysis results, and can decrease the likelihood of maloperation \cite{kundur1994power,sauer1998power,vaiman2012risk,kosterev1999model}. 
%Several IEEE power flow test cases are developed in \cite{IEEECase}, and can be used for both steady-state and dynamic analysis. 
Several IEEE test cases with dynamic models and parameters for sixth-order generator machines are defined in \cite{demetriou2015dynamic, van2015test}. Work \cite{semerow2015dynamic} utilizes the Dynamic Study Model approach to represent the global dynamics of the entire continental Europe power system, but the data is not openly available due to legitimate security concerns. Restricted public availability of actual large-scale power grid cases greatly 
impede the development of dynamic studies. References \cite{birchfield2016statistical,xu2018modeling} give one possible solution by developing fictitious yet realistic large-scale synthetic power system models that are capable of simulating system transients.

With the increasing penetration of renewable energy resources, the electrical power system is exposed to new balancing and control challenges \cite{xu2018metric,erlich2006impact,von2013time,liu2018locational,muljadi2008validation}. Among all renewable energy resources, wind and PV plants have the fastest increases of generation around 2020 \cite{EIAWindSolar}. These changes and challenges raise awareness on the need for properly tuned large-scale power system cases which take into account the dynamic characteristics of the wind and PV plants. As such, this paper tackles the need to include dynamic models for the wind and PV generators in the large-scale synthetic power grids for transient stability studies.

Researchers and industry have developed standard, flexible and publicly available models for wind and solar PV generation technologies in the past two decades \cite{conseil2007modeling,sanchez2015generic,ellis2011description}. The recent development of second-generation generic models for wind and PV plants allows a wider range of control strategies  and equipment representation. Generic models also have the merits of public availability and documentation, good portability, validation and flexibility. Moreover, these  models are suitable for conducting futuristic researches since they have been functionally validated, so that the sensitivity of system response to diverse control strategies can be explored \cite{pourbeik2016generic}. Utilizing these second-generation generic models, this paper focuses on integrating renewable generators in large-scale synthetic network dynamic case for dynamic studies.

Previous work \cite{xu2018modeling,xu2017creation} performed statistical analysis on various machine/exciter/governor/stabilizer models for multiple fuel types, excluding wind and solar generators. This paper aims to further extend the dynamic cases with wind and solar generation models. Statistical analysis on actual power system cases is conducted for wind and solar generators. The composition ratio of different types of wind turbine models is identified and assigned to each bus that has a wind generation unit. Generic modules for generator/excitation/aerodynamics/pitch controller/torque controller/plant controller are used for different type wind plants if applicable \cite{council2014wecc}. Models of generation/excitation/plant controller of central station solar plants are implemented as well \cite{council2019wecc}. Templates of parameters are extracted from actual cases based on various model type and different control strategies. One of multiple parameter templates for each module is assigned to a synthetic generator. In this work, the ACTIVSg2000 case (with 2000 buses) and the ACTIVSg10k case (with 10K buses) \cite{ElectricGrid} are used as illustrations for integrating renewable generators. Simulations results are displayed to show the ability to provide real and reactive power support at the point of interconnection (POI). Several transient stability metrics are used to verify that the case has adequate performances after N-1 contingencies. 
%Example of functionality of a plant controller is given to show that the applied model can control all the active devices in the power plant to achieve a shared objective  at the point of interconnection (POI) with the rest of the power grid. 

The remainder of the paper is organized as follows. In section \RNum{2}, a procedure is developed to automatically identify and assign a model and a parameter template to each renewable generator. Simulation results using the 2000-bus case and the 10K-bus case integrated with renewable generators dynamic models are given in Section \RNum{3} successively. Last, Section \RNum{4} offers conclusions of this paper and future work direction.

\section{Extension of Synthetic Network Dynamic Models with Renewable Generators}
A key aspect of dynamic simulations of power systems is to determine the proper models for different types of generators. An approach of assigning dynamic models and parameters to each renewable generator is developed in this section. In particular, the composition ratio of different types of WTGs is taken from the statistics of actual cases and then mapped to each WTG. Multiple parameter templates are extracted from actual cases. Then suitable models and associated parameter template are joined to obtain renewable generators' diverse  dynamic model profiles. This section starts with determining the composition ratio of WTGs.

\begin{figure}[h]
\setlength{\textfloatsep}{5pt}
\centerline{\includegraphics[scale=0.535]{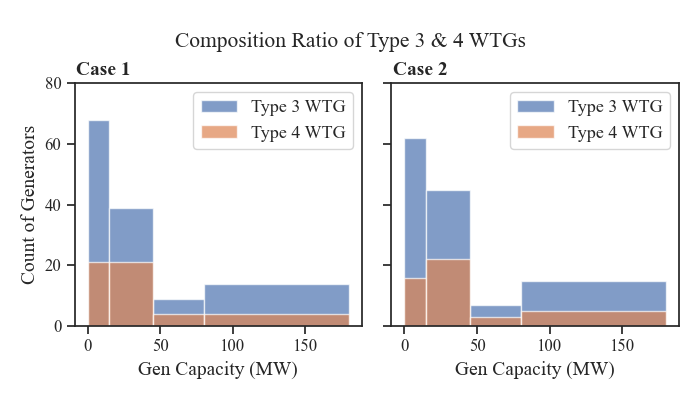}}
\caption{Composition ratio of Type 3 \& 4 WTGs extracted from actual cases}
\label{Comp Ratio}
\end{figure}

Newly installed wind power plants are essentially all of Type 3 or 4 due to the limitations of Type 1 \& 2 WTGs \cite{council2014wecc}. Only the former types are considered in statistics analysis and then assigned to synthetic generators with the same probabilities as their relative proportion. For example, the composition ratio of Type 3 \& 4 WTGs from two  Western  Electricity  Coordinating  Council (WECC) cases are approximately one third and two thirds after computing their relative proportion shown (Fig.\ref{Comp Ratio}). In Fig.\ref{Comp Ratio}, the width of each bar indicates the range of capacity of the WTGs that are counted in this bin. 

In the rest of this section, we will discuss the modular approach of renewable energy modeling and then describe the extraction and assignment process of the parameter template for both the high level and low level control modules.

\subsection{The Modular Approach of Renewable Energy Modeling}
Dynamic representation of the wind and solar generator uses a subset of the seven modules shown in Table \ref{module} that form the basis of renewable energy system for dynamic studies. As shown, some models have several versions that are quite similar to each other, such as two renewable energy plant controller models (repc\_a and repc\_b). The difference between repc\_a and repc\_b is in the amount of renewable generator that is being controlled. If there are multiple renewable generators that are controlled by the coordinated plant controller, the use of repc\_b is required; otherwise, repc\_a can be used. By properly connecting these modules, as shown in Table \ref{mod connection}, one is able to create different renewable energy models. The details and the validation of each module can be found in \cite{council2014wecc,council2019wecc,pourbeik2016generic}.

\begin{table}[t]
\begin{threeparttable}
\caption{The basic modular blocks of the renewable energy system}
\label{module}
\begin{center}
\begin{tabular}{|c|c|}
\hline
 Model Name&Model Description \\
 \hline
\hline
 regc\_a&Renewable Energy Generator/Converter Model \\
 \hline
    \makecell{reec\_a \\ (reec\_b \& reec\_c)}& Renewable Energy Electrical Controls Model\\
\hline
    \makecell{repc\_a \\ (repc\_b)} &Renewable Energy Plant Controller Model\\
\hline
wtga\_a&Wind Turbine Generator Aerodynamics Model\\
\hline
wtgt\_a&Wind Turbine Generator Shaft Model\\
\hline
wtgp\_a&Wind Turbine Generator Pitch Controller Model\\
\hline
wtgq\_a&Wind Turbine Generator Torque Controller Model\\
\hline
\end{tabular}
\begin{tablenotes}
\footnotesize
\item Note: The details of each module are documented in \cite{council2014wecc,council2019wecc}
\end{tablenotes}
\end{center}
\end{threeparttable}
\end{table}

\begin{table}[t]
\caption{Combining Modular Blocks to Modeling Renewable Energy}
\label{mod connection}
\begin{center}
\begin{tabular}{|c|c|}
\hline
 RES&Model Combination \\
 \hline
\hline
 Type 3 WTG&\makecell{regc\_a, reec\_a, repc\_a, wtgt\_a, wtga\_a, wtgp\_a, \\wtgq\_a (optional: repc\_b)}  \\
 \hline
Type 4 WTG& \makecell{regc\_a, reec\_a, repc\_a, wtgt\_a \\ (optional: repc\_b)}\\
\hline
PV plant &\makecell{regc\_a, reec\_a, repc\_a \\ (optional: repc\_b)}\\
\hline
\end{tabular}
\end{center}
\end{table}

\subsection{Parameter Assignment for High Level Control Modules}\label{highlevel}
Once finishing the selection and assignment of the modular blocks, a corresponding parameter template is chosen for each module of every renewable generator. We use Type 3 wind power plants as a representation example as it involves the maximum amount of modular blocks in building the model.

Multiple combinations of high level (i.e., plant-level or inverter-level) real and reactive power control are feasible by setting relevant parameters and switches \cite{council2014wecc}. Table \ref{Control option} summarizes a list of dominant control options and the involved modules and switches that are available in WECC cases. Associated with each control option, relative parameters need to be assigned.
Note that, parameter assignment executed individually for each variable may be overly simplified. One needs to consider the physical relations, characteristic values and proper tuning of these parameters during the parameter determination process. Initially, dominant combinations of all variables are extracted from the WECC cases and saved as different parameter templates for each module of every unit. 
%But the authors will manually tune the system after finishing all model and parameter assignment process. 

These recorded data for both switch status and corresponding parameter templates summarized from actual cases are used to create dynamic case profiles for modules that involved in plant-level or inverter-level controls. Except for these modules, several other modules that describe the dynamics inside the power plant are also included, and we will discuss them in the next subsection.

\begin{table*}[t]
\caption{Control Options and Module/Switch Involved}
\label{Control option}
\begin{center}
\begin{tabular}{|c|c|c|c|c|c|}
\hline 
\multicolumn{6}{|c|}{Reactive Power Control Options} \\
 \hline
\hline
 Behaviour in Response to&Mode of Operation&Required Models&Vflag&Qflag&Refflag  \\
 \hline
 Voltage deviations&Plant level V control&regc\_a, reec\_a, repc\_a (repc\_b)&N/A&0&1\\
\hline
Voltage deviations& \makecell{Plant level Q control +\\local coordinated V/Q control} &regc\_a, reec\_a, repc\_a (repc\_b)&1&1&0\\
\hline
Voltage deviations& \makecell{Plant level V control +\\local coordinated V/Q control} &regc\_a, reec\_a, repc\_a (repc\_b)&1&1&1\\
\hline
\multicolumn{1}{c}{~} \\[-0.8ex]
\hline
\multicolumn{6}{|c|}{Real Power Control Options} \\
 \hline
 \hline
  Behaviour in Response to&Mode of Operation&Required Models&Freqflag&Ddn&Dup  \\
  \hline
 Frequency deviations&\makecell{Governor response with\\down regulation, only}&regc\_a, reec\_a, repc\_a (repc\_b)&1& $>$0&0 \\
 \hline
Frequency deviations&\makecell{Governor response with\\up and down regulation}&regc\_a, reec\_a, repc\_a (repc\_b)&1& $>$0&$>$0\\
\hline
\end{tabular}
\end{center}
\end{table*}

\subsection{Parameter Assignment for Low Level Control Modules}
The modules that engage in low level (i.e., individual components) controls are introduced in this subsection separately.
\subsubsection{Drive-train Model}
\begin{figure}[h]
\centerline{\includegraphics[scale=0.44]{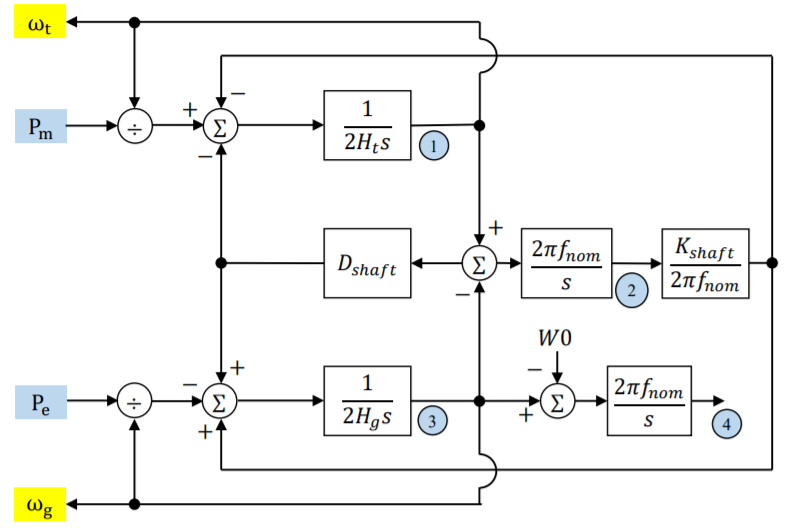}}
\caption{Block diagram for Drive-train Model WTGT\_A}
\label{Drive-train}
\end{figure}

Fig. \ref{Drive-train} shows the block diagram for the governor (drive-train) model WTGT\_A. In these block diagrams, the inputs are colored blue, the autocorrection properties are colored green and the outputs are colored yellow. The parameter assignment process of all low level control modules is similar with high level control modules discussed in Section \ref{highlevel}. Various combinations of parameters are extracted and summarized in Table \ref{wtgt}. Differentiated by the type of governor response limits, there are two general types of parameter templates, with the total count of 9 and 14 respectively. The range of parameters in these templates are also listed for each type of response limits. The next step is to assign a parameter template to each WTG to match the relative probability that these templates appear in the actual data set. 

\begin{table}[h]
\caption{Parameter Templates for WTGT\_A Module}
\label{wtgt}
\begin{center}
\begin{tabular}{|c|c|c|c|c|c|}
\hline
 \makecell{Response\\ Limits} &Ht&Hg&Dshaft&Kshaft&\makecell{Number of\\Templates} \\
 \hline
%\hline
 Fixed& 2.6-6.7&0-5.7&1.01-1.5&-0.08-177&9  \\
 \hline
Normal& 4-7.4&0-0.7&0.6-1.5&-0.08-125&14\\
\hline

\end{tabular}
\end{center}
\end{table}

\subsubsection{Pitch-controller Model}
\begin{figure}[h]
\centerline{\includegraphics[scale=0.46]{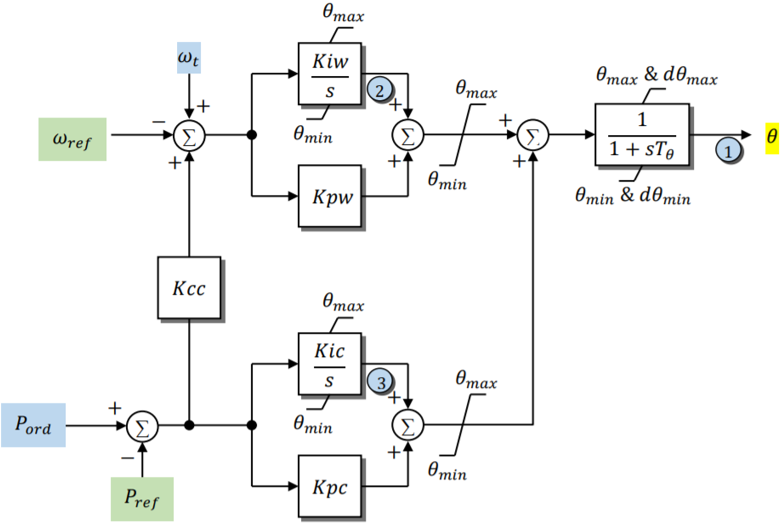}}
\caption{Block diagram for Pitch-controller Model WTGPT\_A}
\label{Pitch}
\end{figure}

Fig. \ref{Pitch} shows the block diagram for the stabilizer (pitch-controller) model. Seven parameter templates are extracted from the actual case, with differences in proportional/integral gain of pitch control/compensator as listed in Table \ref{wtgpt}. \textit{RThetaMin = -10} and  \textit{RThetaMax = 10} are considered since 100\% of modules in the WECC case have \textit{RThetaMin} of -10 and \textit{RThetaMax} of 10. For parameter (\textit{ThetaMin}, \textit{ThetaMax}), values are given (0, 17), (-4, 27) and (0, 27) by probabilities of 0.11, 0.17 and 0.72, respectively. Note, this module can only be used with a Type 3 WTG.

\begin{table}[h]
\caption{Parameter Templates for WTGPT\_A Module}
\label{wtgpt}
\begin{center}
\begin{tabular}{|c|c|c|c|c|c|c|}
\hline
 Tp &Kpp&Kip&Kpc&Kic&Kcc&\makecell{Number of\\Templates}\\
 \hline
%\hline
 0.3&0.7-180& 4-30&0.7-3&4-30&0-0.9&7  \\
 \hline

\end{tabular}
\end{center}
\end{table}

\subsubsection{Torque Model}
\begin{figure}[h]
\centerline{\includegraphics[scale=0.41]{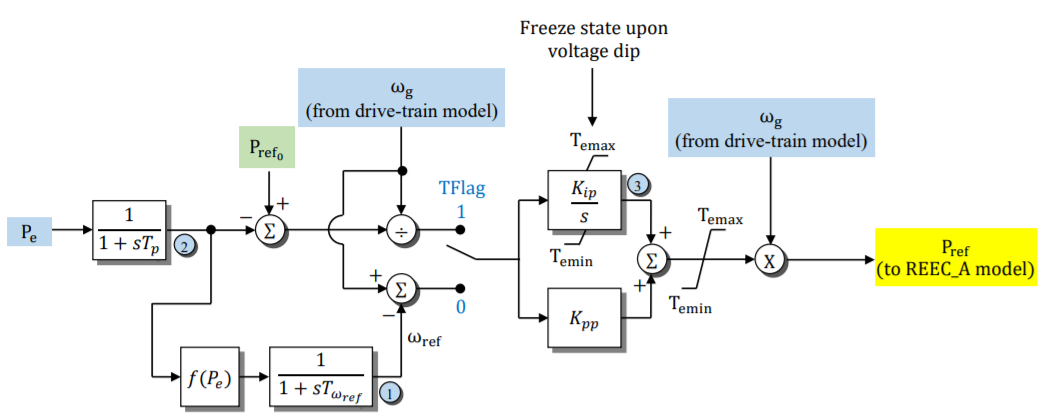}}
\caption{Block diagram for Torque Model WTGTRQ\_A}
\label{Torque}
\end{figure}

Fig. \ref{Torque} shows the block diagram for the Pref Controller (torque) model. As shown in Table \ref{torque}, two general types of templates are extracted from actual cases in regard to the control flag with the total amount of 8 and 1 respectively. Last, \textit{Twref} is set to 60 since 100\% of \textit{Twref} from the WECC case is 60. Different \textit{f(Pe)} functions are also considered in the parameter template, but the details of each function are omitted in this paper due to limitation of pages. Note, this module can only be used with a Type 3 WTG.

\begin{table}[h]
\caption{Parameter Templates for WTGTRQ\_A Module}
\label{torque}
\begin{center}
\begin{tabular}{|c|c|c|c|c|c|c|}
\hline
 \makecell{Control\\ Flag} &Kip&Kpp&Tp&Temax&Temin&\makecell{Number of\\Templates} \\
 \hline
%\hline
 Torque& 0.4-10&2-10&0.02-0.4&1-1.2&0-0.08&8  \\
 \hline
Speed& 0.6&3&0.05&1.2&0.08&1\\
\hline

\end{tabular}
\end{center}
\end{table}

\subsubsection{Aerodynamics Model}
The default data set for this module is used since 100\% of the modules used from the studied cases are utilizing the default values. Also, this module can only be used with a Type 3 WTG.
\section{Illustrative Simulation Results}
In this section, we apply the proposed methodology to model dynamics of renewable generators in two synthetic network cases. We will start with testing aforementioned multiple control strategies by using the 2000-bus case, and provide illustrative simulation results. Then, we will present another example using the 10K-bus case and adopt some transient stability metrics to validate its transient stability performance. The simulation is done in PowerWorld Simulator and a python package, ESA, is used to obtain the simulation results \cite{ESA}. Examples illustrate the renewable generators integrated case have stable dynamic response with well-damped oscillations. 
\subsection{Case \RNum{1} - 2000-bus Case}
\begin{figure}[h!]
\centerline{\includegraphics[scale=0.40]{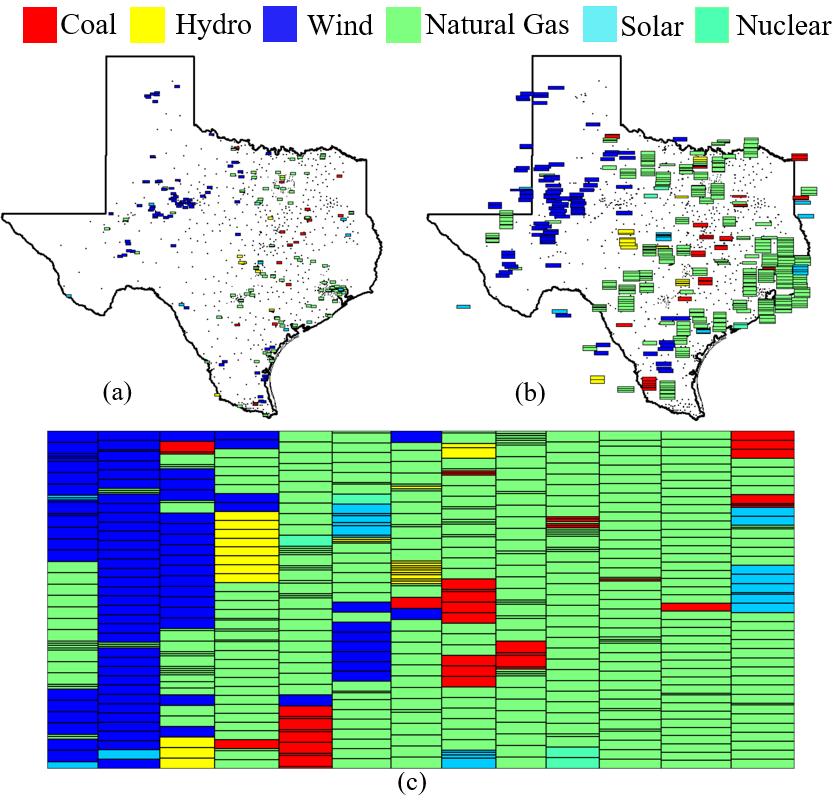}}
\caption{Pseudo-geographic mosaic display of the 2000-bus case \cite{overbye2019wide}\textcolor{white}{Wind, Solarand}with (a) 0\% transitioned (b) 25\% transitioned (c) 100\% transitioned}
\label{Texas_footprint}
\end{figure}

The first case study uses the 2000-bus case \cite{ElectricGrid}, which is built on the Electric  Reliability Council of Texas  (ERCOT) footprint. 
The pseudo-geographic mosaic displays (PGMDs) of this case are shown in Fig. \ref{Texas_footprint}. The row and column locations of the PGMDs approximately represent the object's geographic location \cite{overbye2019wide}, and the size of each rectangular is designed to be proportional to it's real power output in this case. The 544 generators are colored differently according to their fuel types. Examples with varied geographic precision are given with (a) 0\% transitioned from original location, (b) 25\% transitioned and (c) 100\% transitioned. 
% The pseudo-geographic mosaic displays of this case with information about the fuel type of the 544 generators in the system are shown in Fig. \ref{Texas_footprint}, with (a) being 0\% transitioned from original location, (b) 25\% transitioned and (c) 100\% transitioned. 
%Eight areas are defined in this model. 
The 109 renewable generators have a total capacity of about 13 GW. Simulations have verified that this case has a flat start and stable performance in selected N-1 contingencies.
%(loss of a generator or balanced three-phase fault at a bus). 
The integrated renewable generators have been extensively tested and validated for several plant-level or inverter-level control options, as listed in Table \ref{Control option}. Simulation results for a critical N-2 contingency event and a N-1 event are shown as examples in this section. 

\subsubsection{Voltage Deviations at a PV Plant} 
\begin{figure}[h!]
\centerline{\includegraphics[scale=0.46]{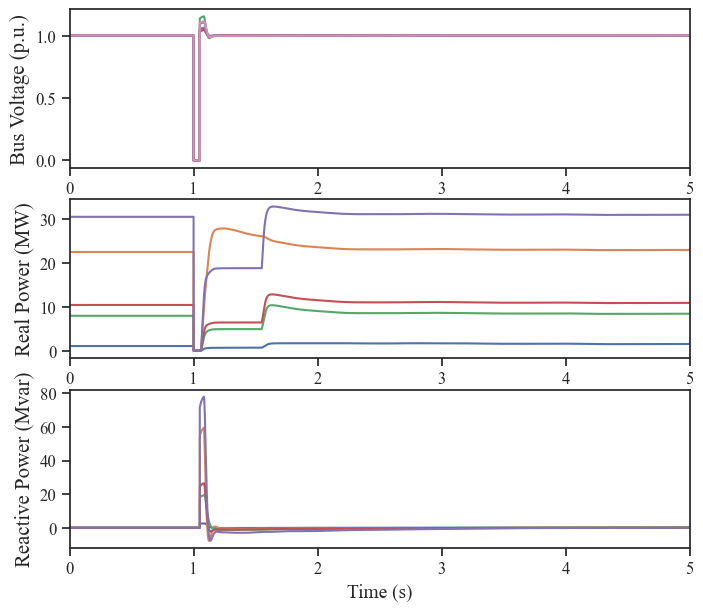}}
\caption{Voltage dip at POI for a duration of 0.05s}
\label{V devia}
\end{figure}

\hfill\\
\label{Reactive Power Support}
Fig. \ref{V devia} shows the simulation results for a PV plant with five individual solar generators experiencing a voltage drop event caused by bus fault at the POI. The event starts at 1.00s and is cleared at 1.05s. The mode of operation of the PV plant is plant level V control plus local coordinated V/Q control. At the beginning of the simulation, the plant controller module REPC\_B is constantly monitoring the bus at POI (i.e., the regulated bus). At 1.00s, the voltage magnitude at the regulated bus falls below the pre-defined criteria and the voltage control function in REEC\_A module is activated. The reactive power output of all the solar generators greatly increased to provide voltage support to the regulated bus. This functionality is vital since during a fault condition the renewable generators should keep connected to reduce the effect of event. Simulation results show this operation mode has impact on the voltage recovery and it's necessary to incorporate this function in renewables. 

\subsubsection{Frequency Increases at a Wind Power Plant} 
\begin{figure}[h!]
\centerline{\includegraphics[scale=0.4522]{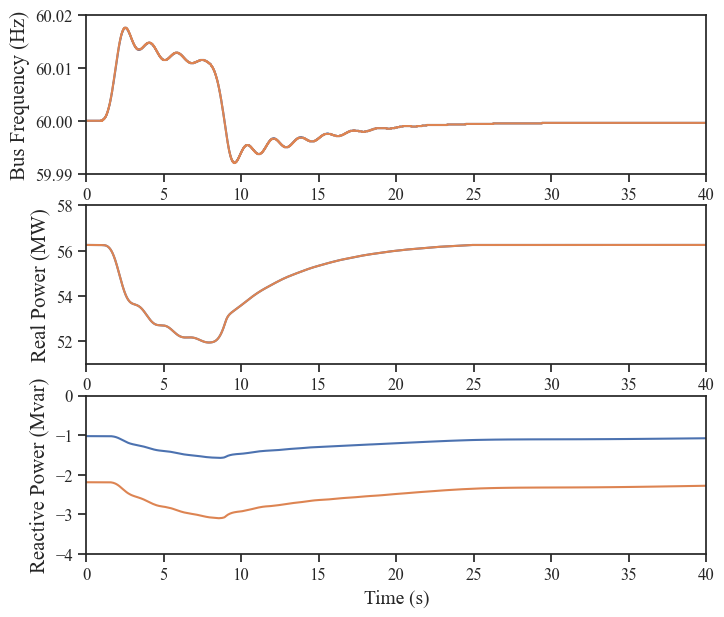}}
\caption{Frequency raise event at POI}
\label{F devia}
\end{figure}

\hfill\\
\label{Real Power Support}
Fig. \ref{F devia} provides the simulation results for a wind plant with two wind turbine generators undergoing an over-frequency event at the POI. The mode of operation at the wind power plant is governor response with down regulation, only (i.e., with no operating reserve for primary frequency response). The event begins at 1.00s, then starts to recover at around 8s and reaches steady-state at around 27s. The plant controller is continuously monitoring the bus at POI. At 1.00s the bus frequency at the regulated bus starts to increase and the frequency control function in REPC\_B module is activated. The real power of all the wind generators in the plant start to ramp down and provide downside frequency support to the system. This feature is of great importance in maintaining the power system frequency's quality and thus it should be included when modeling system dynamics. 
%\begin{figure}[h!]
%\centerline{\includegraphics[scale=0.49]{Frequency Deviation.png}}
%\caption{Frequency drop event at POI}
%\label{F devia}
%\end{figure}

\subsubsection{Frequency Decreases at a Wind Power Plant} 
\begin{figure}[h!]
\centerline{\includegraphics[scale=0.455]{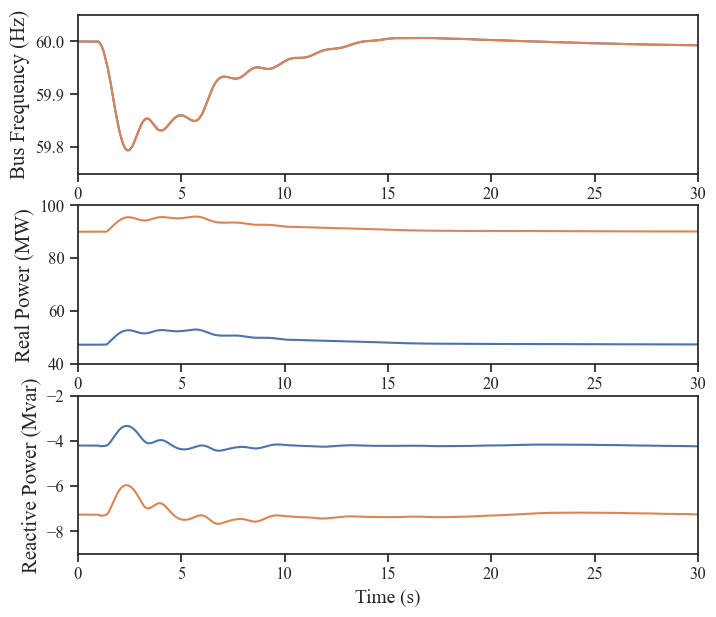}}
\caption{Frequency drop event at POI}
\label{F devia up}
\end{figure}

\hfill\\
Fig. \ref{F devia up} provides the simulation results for a wind plant with two wind turbine generators undergoing a under-frequency event at the POI. The operation mode at the wind power plant is governor response with up and down regulation. The frequency drop event starts at 1.00s, the frequency at POI starts to recover and finally at around 25s return to steady-state. The plant controller is constantly monitoring the regulated bus and the frequency control function is activated when the frequency starts to drop. Then the real power output of the controlled WTGs start to ramp up and provide upward primary frequency response to the system.

Note that providing downward regulation for an over-frequency event is always possible no matter whether or not the renewable resources are operating at the maximum power point. However, only renewable generators that are operating at a sub-optimal power point (i.e., with additional energy headroom to sustain the response) can provide the upward frequency regulation. Operating renewable power plants at a sub-optimal condition to provide upward frequency response have economic consequences while this ability provides a vital reliability benefit \cite{nerc2020guide}. Both aspects are considered in this work. About 15$\%$ of all renewable generators are set to be operating at a sub-optimal condition, which is the same as we have summarized from the actual WECC cases. 

\subsubsection{Illustrative Simulation Results} \hfill\\
After testing the plant-level or converter-level control functionalities of the renewable generators integrated 2000-bus case, this part will focus on demonstrating and validating the dynamic response of the full system. As illustrations, a N-2 contingency event and a bus fault event are considered in the simulation. Case details for both the original 2000-bus case (``Original Case") and the renewable generators integrated case (``Renewable Case") are given in Table. \ref{case details}. The 109 renewable generators are assigned with suitable dynamic models and parameters in the Renewable case, while only power flow data for these generators is available in the Original Case.

\begin{table}[h!]
\caption{Details for the Original Case and the Renewable case}
\label{case details}
\begin{center}
\begin{tabular}{|c|c|c|}
\hline
 &Original Case&Renewable Case \\
 \hline
%\hline
 \makecell{Considered Fuel Type\\in Dynamic Cases} & \makecell{Coal, Natural Gas,\\Nuclear, Hydro}&\makecell{Coal, Natural Gas,\\Nuclear, Hydro\\\textcolor{BlueViolet}{Wind, Solar}}\\
 \hline
$\#$  of Generators& 544&544\\
\hline
$\#$  of Machine Models& 435&544\\
\hline
\multicolumn{1}{c}{~} \\[-1.0ex]
\hline
\multicolumn{3}{|c|}{Changes in the Renewable Case} \\
 \hline
\makecell{$\#$ of Added Dynamic\\ Models by Fuel Type}&\makecell{\makecell{\textcolor{BlueViolet}{Wind}\\\textcolor{white}{\scriptsize{Wind}}}\\\textcolor{BlueViolet}{Solar}}&\makecell{\makecell{58 + 29 = 87\\\scriptsize{(Type 3 + Type 4)}}\\ 22}\\
\hline
\multicolumn{1}{c}{~} \\[-1.1ex]
\hline
\multicolumn{2}{|c|}{$\#$ of Added Machine / Exciter Models}& 109\\
\hline
%\multicolumn{2}{|c|}{$\#$ of Added Exciter Models}& 109 \scriptsize{(REEC\_A)}\\
%\hline
\multicolumn{2}{|c|}{$\#$ of Added Governor Models}& 87 \\
\hline
\multicolumn{2}{|c|}{$\#$ of Added Stabilizer / Aerodynamic Models}& 58\\
\hline
%\multicolumn{2}{|c|}{$\#$ of Added Aerodynamic Model}& 58 \scriptsize{(WTGAR\_A)}\\
%\hline
\multicolumn{2}{|c|}{$\#$ of Added Plant Controller Models}& \makecell{70 + 13\\ \scriptsize{(REPC$\_$A + REPC$\_$B)}}\\
\hline
\multicolumn{2}{|c|}{$\#$ of Added Pref Controller Models}&58 \\
\hline
\end{tabular}
\end{center}
\end{table}

\begin{figure}[h]
\centerline{\includegraphics[scale=0.5]{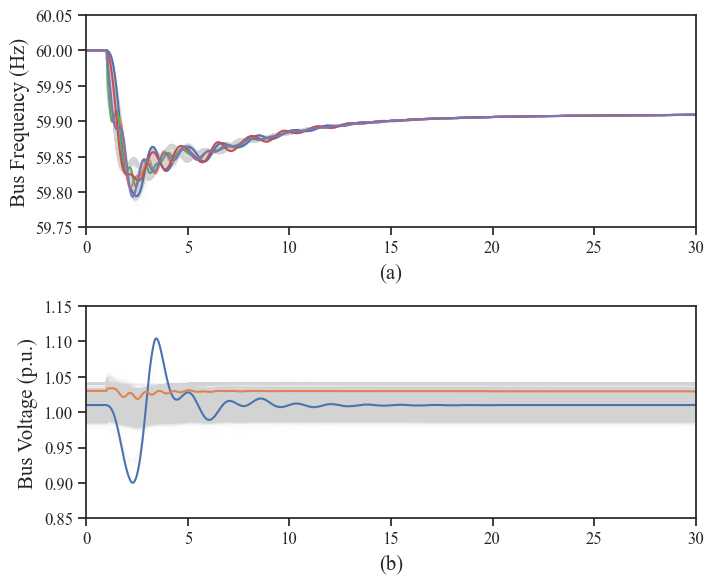}}
\caption{Simulation results using the Renewable Case after a N-2 contingency event (Solid color lines for representative bus frequency (voltage) profiles; Light-grey lines for all bus frequencies (voltages))}
\label{2000_GenDrop}
\end{figure}

The first example is a N-2 contingency event. Two large generators with total real power output of 2589 MW are disconnected from the system at 1.00 s. The simulated bus frequencies and voltages collected from the Renewable Case are displayed in Fig. \ref{2000_GenDrop} in light-grey lines. Some representative values are plotted in solid color lines.
%In order to assess the stability level of the Renewable Case after the severe N-2 contingency event, modes for frequency and voltage signals are calculated using the Iterative Matrix Pencil (IMP) method introduced in \cite{trinh2019iterative}. Oscillations with a minimum frequency of 0.4Hz \cite{england2019transmission} and a maximum frequency of 2Hz \cite{pierre1997initial} are considered in this work. 
%In each mode, the signal with the largest component is plotted in solid color lines in Fig. \ref{2000_GenDrop}. 
We observe that the bus frequencies (voltages) start to recover soon after the contingency event and settle down within 20s. 

\begin{figure}[h!]
\centerline{\includegraphics[scale=0.53]{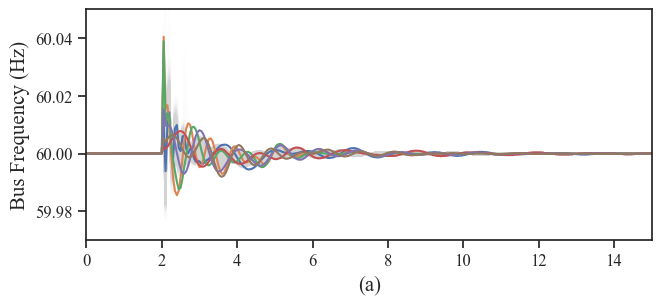}}
\caption{Simulation results using the Renewable Case for a three-phase fault on a 115kV bus (Solid color lines for representative bus frequency profiles; Light-grey lines for all bus frequencies)}
\label{2000_BusFault}
\end{figure}

The second example is a bus fault event. A three-phase fault is applied to a 115kV bus at 2.00s and cleared at 2.05s. Simulation results for frequencies of all buses are plotted using light-grey lines in Fig. \ref{2000_BusFault}. Similarly, the solid color lines represent some representative bus frequency values. The fast dynamics in bus frequencies are eliminated within seconds after the fault is cleared as shown in the results. The bus voltages drop immediately when the fault applies and they begin to recover at the fault clearing time. The post-fault voltages amplitudes of all buses are approaching to their pre-faut values. 

%\vspace{-2.9mm}
%Fig. \ref{2000_GenDrop} (Fig. \ref{2000_BusFault}) displays bus frequency and voltage profiles after the system is subject to a N-2 contingency event. Simulation results verify that both the Original Case and the Renewable Case have flat start. As shown in Fig. \ref{Original_Case} (a) and Fig. \ref{New_Case} (a), the bus frequencies drop instantly as the contingency event happens, then start to recover when frequencies crosses the nadir. Bus the value of post-contingency frequency nadir of the Renewable Case (59.775) is higher than that of the Original Case (59.732). As discussed in Section \ref{Real Power Support}, the added renewable generators dynamic models have the capability of providing real power support to the system and result in a higher frequency nadir and a faster frequency response. Simulation results for bus voltages are given in Fig. \ref{Original_Case} (b) and Fig. \ref{New_Case} (b). Voltage magnitudes start to change at 1.00 s and the boundary envelope of voltage profiles are marked as maroon dotted lines in the figures. We observe that the voltage boundary envelope of the Renewable Case is narrower than that of the Original Case. As discussed in Section \ref{Reactive Power Support}, including renewable generators' dynamic models have impact on system voltage stability and thus this process is crucial for dynamics studies. Therefore, simulation results prove that proper modeling of dynamic models for renewable resources can be helpful for maintaining system stability. 
\subsection{Case \RNum{2} - 10K-bus Case}
\begin{figure}[h!]
\centerline{\includegraphics[scale=0.46]{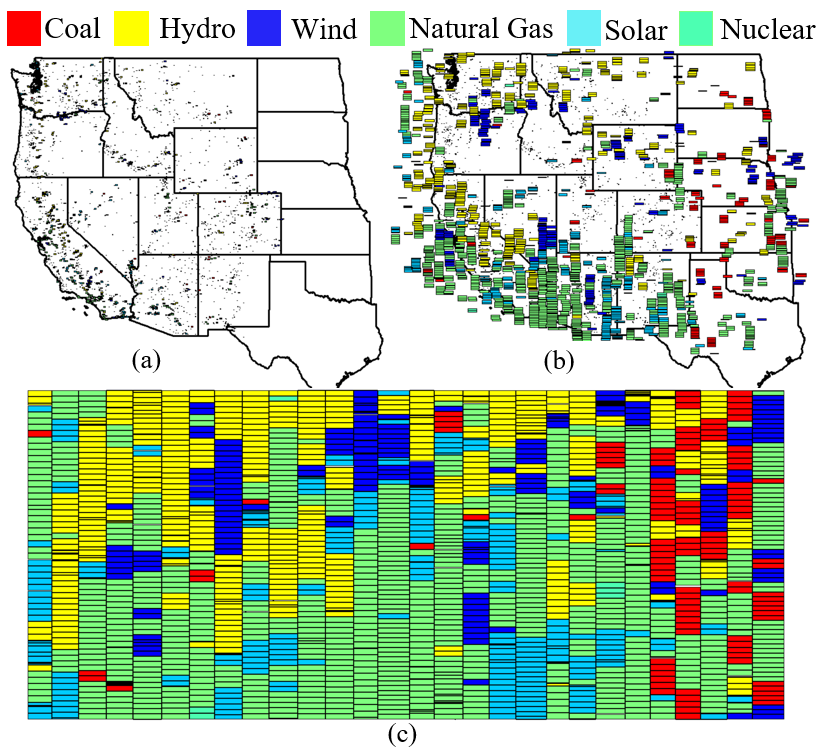}}
\caption{Pseudo-geographic mosaic display of the 10K-bus case \cite{overbye2019wide}\textcolor{white}{Wind, Solarand}with (a) 0\% transitioned (b) 25\% transitioned (c) 100\% transitioned}
\label{10k_footprint}
\end{figure}

The second case study is using the 10K-bus test case built on footprint of western United States, which is available at \cite{ElectricGrid}. The pseudo-geographic mosaic displays of the case are given in Fig. \ref{10k_footprint}. Sixteen areas and seven nominal voltage levels are defined. The 634 renewable generators in this case have a total capacity of about 32 GW. First, we run simulation for 100s without any contingency or fault event to verify that this case has a flat start. Then, selected N-1 contingencies (generator drop and three-phase bus fault) are applied to disturb the system. For each event, we run transient stability analysis and further analyze these simulation results.  

\subsubsection{Transient Stability Validation Metrics} \hfill\\
Reference \cite{xu2018modeling} considers three transient stability metrics for N-1 contingencies: the minimum damping ratio of generator rotor angle (\textit{$M_r$}), the minimum and maximum bus frequencies after the last contingency event (\textit{$M_f$}) and the minimum ratio of bus voltage nadir to pre-contingency value (\textit{$M_v$}). The transient stability requirements of these metrics are \textit{$M_r$} larger than 3$\%$, \textit{$M_f$} between [59.5, 60.5] Hz and \textit{$M_v$} larger than 75$\%$.

%\begin{table}[h]
%\caption{Transient Stability Metrics and Requirements}
%\label{TS metrics}
%\begin{center}
%\begin{tabular}{|c||c|c|c|}
%\hline
%Metric&$M_r$&$M_f$&$M_v$ \\
% \hline
%Requirement&3$\%$ & [59.5, 60.5]Hz&75$\%$ \\
%\hline
%\end{tabular}
%\end{center}
%\end{table}
%where \textit{$M_r$} specifies the minimum damping ratio of generator rotor angle; \textit{$M_f$} puts limits on the minimum and maximum bus frequencies after the last contingency event and \textit{$M_v$} defines the minimum ratio of bus voltage nadir to pre-contingency value. 
\begin{figure}[h]
\centerline{\includegraphics[scale=0.4]{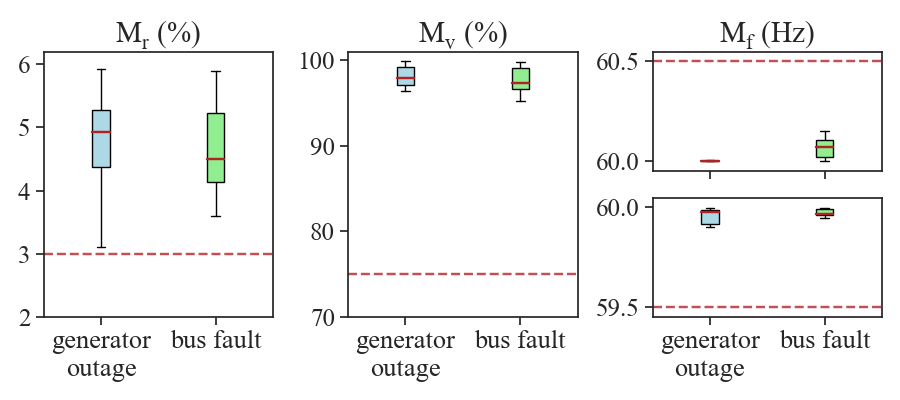}}
\caption{Calculated transient stability metrics using the Renewable Case after being subjected to selected N-1 contingency events}
\label{10k_Metrics}
\end{figure}
Fig. \ref{10k_Metrics} depicts the calculations of these metrics. In the boxplot, the blue box indicates the results of a generator outage event and the green box indicates the results of a bus fault event. The upper and lower extreme lines shows the maximum and minimum value. A box is created from the first quartile to the third quartile of each metric and the solid maroon line which goes through the box indicates the median. These results verify that the Renewable Case meets the transient stability requirements after being subjected to selected contingencies.
\vspace{2mm} %2mm vertical space
\subsubsection{Illustrative Simulation Results} 
\begin{figure}[h]
\centerline{\includegraphics[scale=0.52]{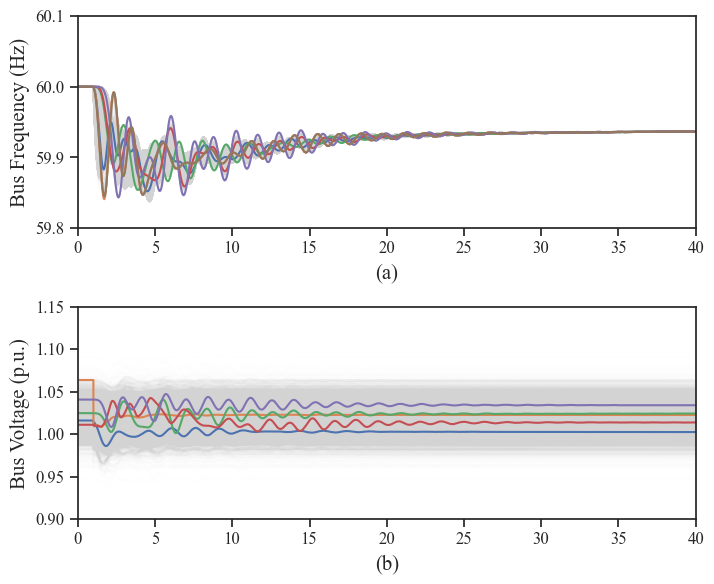}}
\caption{Simulation results using the renewables integrated 10K-bus case after a N-2 contingency event (Solid color lines for representative bus frequency (voltage) profiles; Light-grey lines for all bus frequencies (voltages))}
\label{10k_GenDrop}
\end{figure}

\hfill\\
Simulation results for a N-2 and a N-1 contingency event are given as examples. The results for the N-2 event (loss of two large generators at 1.00s) are given in Fig. \ref{10k_GenDrop}. The frequencies and voltages for all buses are plotted with light-grey lines and selected illustrative signals are highlighted in solid color lines. We can see that bus frequencies and voltages start to vary and oscillate when the N-2 event happens at 1.00s, but the oscillations are damped out within 30s. The post-contingency frequencies and voltages are at acceptable levels for this severe and rare event.

\begin{figure}[h]
\centerline{\includegraphics[scale=0.51]{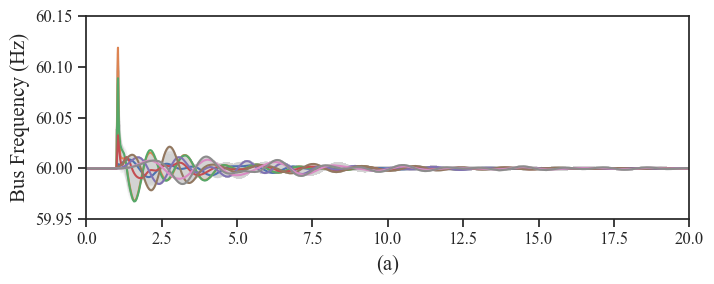}}
\caption{Simulation results using the renewables integrated 10K-bus case after a N-1 contingency event (Solid color lines for representative bus frequency profiles; Light-grey lines for all bus frequencies)}
\label{10k_BusFault}
\end{figure}

Fig. \ref{10k_BusFault} displays the simulation results for a N-1 contingency event. A balanced three-phase fault is applied to a 161kV bus at 1.00s and then cleared at 1.05s. The light-grey area indicates the results for all bus frequencies. Similarly, solid color lines denotes signals that are representative. After the fault is cleared, the frequency variations decline gradually and the system enters an equilibrium condition. We notice that this renewable generators integrated case has satisfactory dynamic responses and meets transient stability requirements. Also, this case performs to have satisfactory system responses and transient simulation results for a severe N-2 event.

%\clearpage
In general, the proposed approach can be used to model renewable generators in large-scale network models. The generated case can be used in many power system studies. One application is to use this renewable generation integrated case as the base for interactive simulations developed in \cite{wallison2020design} and evaluating the participant's performance using metrics described in \cite{liu2020evaluation}.
\section{Conclusion}
This paper has presented an approach for integrating renewable generators with large-scale synthetic grid model for transient stability studies. The authors performed statistical analysis on modular blocks, which are used to represent dynamic models of renewable energy systems. The composition ratio of Type 3 and Type 4 WTGs are summarized from the actual cases and assigned to each wind generator. Multiple control strategies are considered and corresponding parameter templates are extracted from actual cases. Simulation results show that the tested case has a flat start and has good real and reactive power support capability. Illustrative examples are given to verify that the renewable generators integrated cases have good transient simulation performances and they meet transient stability validation metrics. The created case can be used for many kinds of studies, including inertia displacement, modeling and planning of future power systems and sensitivity analysis of system performance to different control strategies, etc. The proposed method is general enough to be applied in modeling the dynamics of any type of generators in any systems. 

%Nevertheless, further tuning and validating of the renewable generators are quite possible. The authors will report the updates of this work in the future. 
\section*{Acknowledgment}
This work was supported by the National Science Foundation under Award Number \textcolor{black}{ECCS-1916142} and the U.S. Department of Energy (DOE) under award DE-OE0000895.

\vspace{12pt}
{
\small
\bibliography{citation}
\bibliographystyle{IEEEtranN}
}

\end{document}